\documentclass[a4paper]{article}

\newcommand{\rem}[1]{\ifodd\value{page} \reversemarginpar
\else \normalmarginpar \fi \marginpar{{\footnotesize #1}} }

\usepackage[centertags]{amsmath}
\usepackage{amsfonts}
\usepackage{amssymb}

\usepackage{graphicx}
\usepackage{rotating}
\usepackage{booktabs}
\usepackage{multirow}
\usepackage[table]{xcolor}
\usepackage{tikz}
\usepackage{caption}

\newcommand{\ba}{\begin{array}}
\newcommand{\ea}{\end{array}}

\newcommand{\be}{\begin{equation}}
\newcommand{\ee}{\end{equation}}

\newcommand{\dis}{\displaystyle}

\newtheorem{rema}{Remark}[section]
\newcommand{\br}{\begin{rema}}
\newcommand{\er}{\end{rema}}

\newtheorem{thm}{Theorem}[section]
\newtheorem{prop}[thm]{Proposition}

\def\mref#1{(\ref{#1})}
\def\eqref#1{(\ref{#1})}
\usepackage{texdraw}
\usepackage{fancyhdr}
\begin{document}

\title{Systems of difference equations on a vector valued function that admit 3D space of scalar potentials}

\author{Pavlos Kassotakis \thanks{\emph{Present address:} Department of Mathematics and Statistics, University of Cyprus, P.O Box: 20537, 1678 Nicosia, Cyprus;
\newline \emph{e-mails:} {\tt pavlos1978@gmail.com, pkasso01@ucy.ac.cy}}
\and Maciej Nieszporski \thanks{\emph{Present address:}Katedra Metod Matematycznych Fizyki, Wydzia\l{} Fizyki, Uniwersytet Warszawski,
ul. Pasteura 5, 02-093 Warszawa, Poland;
\newline \emph{e-mails:} {\tt maciejun@fuw.edu.pl}}
}

\maketitle
\begin{abstract}
For some involutive maps  $\Phi:{\mathbb C}P^1 \times {\mathbb C}P^1 \to {\mathbb C}P^1 \times {\mathbb C}P^1$ we find all invariants with separated variables.
We investigate a link of the maps and their invariants with separated variables to discrete integrable systems.
Maps correspond to integrable systems on edges (bond systems), while their invariants with separated
variables yields potentials of the bond systems, that allows us to rewrite the integrable sytems as
 models on vertices. Among the latter ones one can find well known integrable difference equations as well as difference relations, which in contrast to the equations give non-single-valued evolution of the dependent
variable. However, the non-single-valuedness can be resolved by the link with the bond system.
\end{abstract}


\section{Introduction} \label{Section0}
 Let us consider
a map defined on the cartesian product of  complex projective lines, $\Phi: {\mathbb C}P^1 \times {\mathbb C}P^1 \ni (u, v) \mapsto (U, V) \in {\mathbb C}P^1 \times {\mathbb C}P^1$ where
\begin{equation} \label{eq1.1}
 \Phi: \qquad U={\dis v + \frac{p-q}{u-v}}, \qquad V={\dis u + \frac{p-q}{u-v}}
\end{equation}
and where $p$ and $q$ are some parameters. The map $\Phi$ is an involution. Involutive maps admit a plethora of invariants, where by an {\it invariant} {\it(alternating invariant)} of a map we understand here a function $I$ such that $I(U,V)=I(u,v)$ ($I(U,V)=-I(u,v)$).  Among the  invariants (alternating invariants), there might exist some with separated variables.
  For example, the functions
\begin{equation} \label{eq1.2}
I(u,v)= a (u-v)  + b (u^2+p-v^2-q)+c (u^3+3pu-v^3-3qv)+d,
\end{equation}
where $a$, $b$, $c$ and $d$ are arbitrary constants, are alternating invariants with separated variables for the map~$\Phi.$
Not only that, it turns out that this family of alternating invariants of the map $\Phi$ is complete i.e. no other invariants of the map in separated form  exist. Therefore, the alternating invariants with separated variables of the map $\Phi$ form a 3-dimensional affine space.

\begin{figure}[h] \label{Figure1}
\begin{minipage}[h]{0.5\textwidth}
\begin{tikzpicture}[scale=2, every node/.style={transform shape}];
\coordinate (A) at (0,0);
\coordinate (B) at (2,0);
\coordinate (C) at (2,2);
\coordinate (D) at (0,2);
\draw (A)--(B)--(C)--(D)--(A);
\draw [-latex] (A) -- (C) node [scale=0.5, midway, above] {$\Phi$};
\draw (A)--(B) node [scale=0.8, midway, below] {$u$};
\draw (A)--(D) node [scale=0.8, midway, left] {$v$};
\draw (C)--(D) node [scale=0.8, midway, above] {$U$};
\draw (B)--(C) node [scale=0.8, midway, right] {$V$};
\end{tikzpicture}
\captionsetup{font=footnotesize}
\captionof*{figure}{(a)}
\end{minipage}
\begin{minipage}[h]{0.5\textwidth}
\begin{tikzpicture}[scale=2, every node/.style={transform shape}];
\coordinate (A) at (0,0);
\coordinate (B) at (2,0);
\coordinate (C) at (2,2);
\coordinate (D) at (0,2);
\draw (A)--(B)--(C)--(D)--(A);
\draw (A)--(B) node [scale=0.8, midway, below] {$u_{m_1,n_1}$};
\draw (A)--(D) node [scale=0.8, midway, left] {$v_{m_2,n_2}$};
\draw (C)--(D) node [scale=0.8, midway, above] {$u_{m_1,n_1+1}$};
\draw (B)--(C) node [scale=0.8, midway, right] {$v_{m_2+1,n_2}$};
\end{tikzpicture}
\captionsetup{font=footnotesize}
\captionof*{figure}{(b)}
\end{minipage}
\caption{ (a): Values and arguments of the map $\Phi$. (b): Variables on edges on an elementary square of the $\mathbb{Z}^2$ lattice }\label{notation0}
\end{figure}
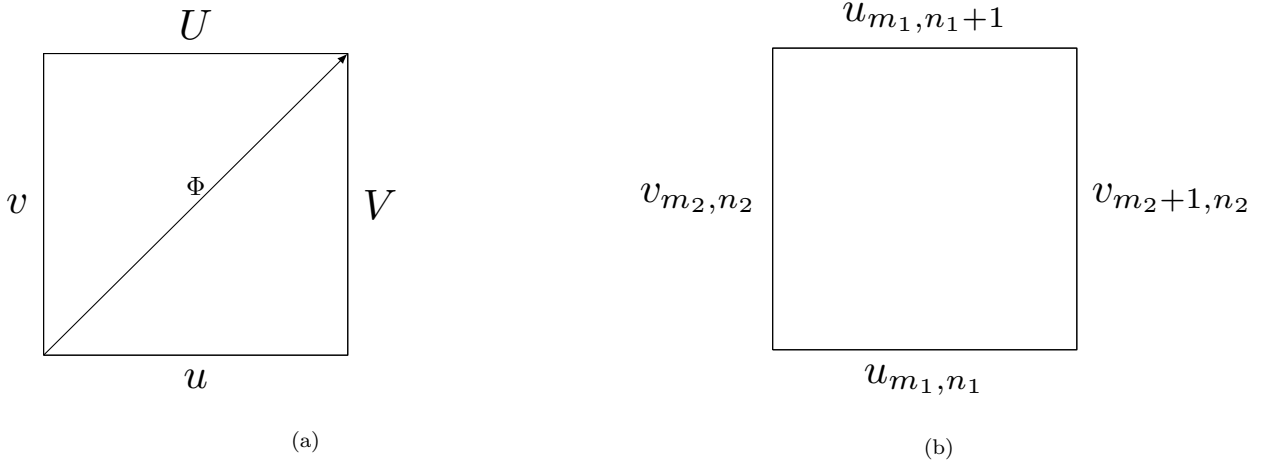

A connection of such   maps with integrable difference equations was pointed out  by several authors
\cite{Korepanov-1998,Sergeev-1998,ABS,Tasos,KaNie,KaNie:2011,KaNie:2018}.
For example having in  hand map  (\ref{eq1.1}) we can define a difference system in two  variables $u$ and $v$ defined on the edges of a ${\mathbb Z}^2$ graph.
We consider $u$ as a function  defined  on the set of {\em horizontal edges}
(i.e. pairs of vertices $\{((m,n),(m+1,n)) \, | \, (m,n) \in {\mathbb Z}^2 \}$)
and respectively $v$ as a function  defined  on set of {\em vertical edges} (i.e. pairs of vertices $(m_2,n_2):=\{(m,n),(m,n+1)\}, (m,n)  \in {\mathbb Z}^2$),  see  Figure \ref{notation0}.

We label the horizontal edges by pair of numbers  $(m_1,n_1)$ such that $((m,n),(m+1,n))\mapsto (m_1,n_1)=(m+\frac{1}{2},n)$ and the vertical edges by pair of numbers  $(m_2,n_2)$ such that $((m,n+1),(m,n))\mapsto (m_2,n_2)=(m,n+\frac{1}{2})$.
Now, taking map (\ref{eq1.1}) we define a difference equation via
\[\forall \; (m,n) \in {\mathbb Z}^2 \, : \quad (u_{m_1,n_1+1},v_{m_2+1,n_2})=\Phi(u_{m_1,n_1},v_{m_2,n_2}), \]
which explicitly reads
\begin{equation} \label{eq1.3}
u_{m_1,n_1+1}={\dis v_{m_2,n_2} + \frac{p_{m_1}-q_{n_2}}{u_{m_1,n_1}-v_{m_2,n_2}}}, \qquad v_{m_2+1,n_2}={\dis u_{m_1,n_1} + \frac{p_{m_1}-q_{n_2}}{u_{m_1,n_1}-v_{m_2,n_2}}},
\end{equation}
where we allow $p_{m_1}$ to be a given function of the independent variable $m_1$ only  and $q_{n_2}$ to be a given function of the independent variable $n_2$ only.  The set of equations (\ref{eq1.3}) are equations on vector valued  two component  function $(u,v)$ that  can be regarded as  field equations. Its potential versions arise from
the invariance condition $I(u_{m_1,n_1+1},v_{m_2+1,n_2})=-I(u_{m_1,n_1},v_{m_2,n_2}).$  Choosing the constants as $a=1, b=c=d=0,$ the invariance condition reads
\begin{equation} \label{eq1.4}
u_{m_1,n_1+1}+u_{m_1,n_1}=v_{m_2+1,n_2}+v_{m_2,n_2}.
\end{equation}
The relation $(\ref{eq1.4})$ guarantees the existence of a potential  $\psi_{m,n}$ defined on vertices of the $\mathbb{Z}^2$ graph such that
$$
u_{m_1,n_1}=\psi_{m+1,n}+\psi_{m,n},\quad v_{m_2,n_2}=\psi_{m,n+1}+\psi_{m,n}.
$$
In terms of the potential function $\psi,$ the difference system $(\ref{eq1.3})$ becomes the lattice potential KdV equation \cite{nij-qui-cap,Nijhoff-1995}, namely
\begin{equation} \label{lpkdv}
(\psi_{m+1,n+1}-\psi_{m,n}) (\psi_{m+1,n}-\psi_{m,n+1})=p_{m_1}-q_{n_2}.
\end{equation}
A question arises, what about other choices of the parameters $a,b,c$ and $d$?
 A partial answer to this  question appeared in our articles \cite{KaNie,KaNie:2011,JJJ,AtkNie,KaNie:2018}, where we were motivated mainly by the fact that nonlinear superposition principle for Korteweg de Vries equation can be reinterpreted as
multi-quadratic relation that arises from the choice of the parameters $b=1,\, a=c=d=0$ in the equations above \cite{AdVeQ,KaNie}.
Here we answer this question in a systematic manner and under some assumptions exhaustively,
revealing the richness of the world of difference relations.

 We start the paper with   the presentation of maps that
have mainly  their origin in the papers \cite{ABSf,PSTV,Tasos}, where they are  referred to as Yang-Baxter maps (see \cite{VeselovYB} and references therein), and we also consider involutive maps which  do not satisfy the Yang-Baxter property (see subsection \ref{subsec2.1}). We provide the complete set of invariants  with separated variables of the mentioned maps, see Sections \ref{Section2}, \ref{Section3} and \ref{Section4}.  Then we reinterpret the results as equations on a lattice in Section \ref{Section5} and we give an overview of vertex equations and relations in Section \ref{Section6}.
We end this article with discussions
  and some propositions for further development in Section \ref{Section7}.


\section{Involutive maps} \label{Section2}

We consider the {\it involutive} maps $\Phi: {\mathbb C} P^1 \times {\mathbb C} P^1 \ni (u,v) \mapsto (U,V) \in {\mathbb C} P^1 \times {\mathbb C} P^1$
of the form
\begin{equation}
\label{M}
U=\frac{\alpha^1(v) u + \alpha^2(v)}{\alpha^3(v) u + \alpha^4(v)}, \qquad V=\frac{\alpha^5(u) v + \alpha^6(u)}{\alpha^7(u) v + \alpha^8(u)},
\end{equation}
where the functions $\alpha^i$, $i=1,\ldots, 8$ are polynomials of the indicated variables.

Adler et  al. \cite{ABSf} considered this kind of maps in their quest of Yang-Baxter maps.
Namely, they investigated maps with polynomials $\alpha^i$ up to the second degree and which are quadrirational\footnote{The notion of {\it $2^n-$rational maps} was introduced in \cite{KaNie:2017} and serves as an extension to  arbitrary dimensions of the notion of quadrirational maps}  i.e.
such maps $(u,v) \mapsto (U,V)$  that all of the maps $(u,v) \mapsto (U,V)$, $(u,V) \mapsto  (U,v)$, $(U,v) \mapsto  (u,V)$, $(U,V) \mapsto  (u,v)$
are rational. They classified such  maps modulo (M\"ob)$^4$-action i.e. up to transformation $(u,v,U,V)\mapsto (u',v',U',V')=(f^1(u),f^2(v),f^3(U),f^4(V))$
where $f^i$, $i=1,2,3,4$  are linear fractional functions of the indicated variable.
Soon later Papageorgiou et al. \cite{PSTV}
argued that in the context of Yang-Baxter maps symmetry group should be reduced  to $(u,v,U,V)\mapsto (u',v',U',V')=(f^1(u),f^2(v),f^1(U),f^2(V))$ which they refer to as (M\"ob)$^2$-action.
In article \cite{Tasos} the authors
investigated the relationship of  integrable lattice equations to these maps. 

We are going to investigate  the involutive maps in the case that the  polynomials $\alpha^i$ are of first degree and we take advantage of the results in \cite{ABSf,PSTV}, where the polynomials are of second degree.
Note that, due to assumed form (\ref{M}) the companions of the map exist but not necessarily the inverse of the map. The involutivity guarantees the existence of the inverse of the map,  so our maps are quadrirational.  Note also that the (M\"ob)$^2$-action preserves the involutivity of the map.

\subsection{Degree 1 case} \label{subsec2.1}
Here we investigate the case where the  polynomials $\alpha^i$  of (\ref{M}) are at most of degree one.
\begin{prop} \label{prop2.1}
Mapping (\ref{M}) with $\alpha^3(v)$ and $\alpha^7(u)$ first degree polynomials and with the remaining $\alpha^i$ polynomials   of degree   one at most, is involutive iff it is of the following form:
\begin{equation} \label{M1}
U=\frac{\widetilde{C} uv+P u+\widehat{L} A v - A G}{uv+\widetilde{K} (u-\widetilde{C}) +\widetilde{L} ( v -\widehat{C})-\widetilde{C}\widehat{C} }, \quad
V=\frac{\widehat{C} uv+Q v+\widetilde{K} B u - B G}{uv+\widehat{K} (u-\widetilde{C}) +\widehat{L} ( v -\widehat{C})-\widetilde{C}\widehat{C}},
\end{equation}
where
\begin{equation}
\begin{array}{c}
A=\widetilde{C}-\widetilde{L}+\widehat{L}, \quad B=\widehat{C}-\widehat{K}+\widetilde{K}, \quad
P=\widetilde{C}\widehat{K}+\widetilde{L}\widehat{K}-\widetilde{L}\widetilde{K}, \quad Q=\widetilde{L}\widehat{C}+\widetilde{L}\widehat{K}-\widehat{L}\widehat{K}, \\
 G = \widetilde{C} \widehat{K} + \widetilde{L} \widehat{K} + \widetilde{L} \widehat{C} - \widehat{L} \widetilde{K} + \widetilde{C} \widehat{C}.
\end{array}
\end{equation}
The   position of the singularities of the map  (\ref{M1})  are:
\[ (\frac{X}{\widetilde{K}+\widehat{C}} + A , B),  \quad
 (\frac{-X}{\widehat{K}-\widetilde{K}} +A , \frac{-X}{\widetilde{L}-\widehat{L}}+B),\]
 for the first part of the map and
\[
 (A ,  \frac{X}{\widetilde{C}+\widehat{L}} +B), \quad
 (\frac{-X}{\widehat{K}-\widetilde{K}} +A , \frac{-X}{\widetilde{L}-\widehat{L}}+B),\]
for the second part of the map, where
\[X=(\widetilde{L}-\widehat{L})(\widehat{C}+\widetilde{K})+(\widehat{L}+\widetilde{C})(\widehat{K}-\widetilde{K}).\]
\end{prop}

The following remarks are in order. First, setting $X=0,$ coalescence of all singularities   take place and the map (\ref{M1}) is (M\"ob)$^2$ equivalent to the following involutive linear map \cite{Franks-book,Kouloukas-2017,Dimakis2017,KaNie:2018}
$$
\qquad u_2={\dis v + \frac{p+q}{p-q}(v-u)},                     \quad v_1={\dis u + \frac{p+q}{p-q}(v-u)}. \hskip \textwidth minus  \textwidth (L_a).
$$
Second, the map  (\ref{M1}) is (M\"ob)$^2$ equivalent to the map
\be \label{hkdv1}
U=v+\lambda (1-v/u), \quad V=u+\lambda (u/v-1).
\ee
or to the map
\be \label{c-hirota-map-auto1}
U=\frac{\lambda v}{\lambda+(v-u)} ,\quad  V=\frac{\lambda u}{\lambda+(v-u)}.
\ee
Note that the parameter $\lambda$ can be set to one (scaled off) but we left it for historical reasons.

Mapping (\ref{hkdv1}) is the Hirota's KdV map  which is closely related to Hirota's KdV equation \cite{KaNie:2011}, while the map (\ref{c-hirota-map-auto1}) is the companion map (the companion of a map $(u,v)\mapsto (U,V)$  is the map $(u,V)\mapsto (U,v)$) of (\ref{hkdv1}).
The singularities of   map  (\ref{hkdv1}) are  $(0,0)$, $(\lambda,\infty)$ for the first part of the map and
$(0,0)$, $(\infty,-\lambda)$ for the second part. Whereas the singularities of  map  (\ref{c-hirota-map-auto1}) are  $(\infty,\infty)$, $(\lambda,0)$ for the first part of the map and  $(\infty,\infty)$, $(0,-\lambda)$ for the second part.

\subsection{Degree 2} \label{ss22}

The starting point of our considerations   is the F-list of quadrirational Yang-Baxter maps \newline
$\mathbb{CP}^1 \times \mathbb{CP}^1 \ni (u,v)\mapsto (U,V) \in \mathbb{CP}^1\times \mathbb{CP}^1 $ given in \cite{ABSf}
\begin{displaymath}
\label{F}
\begin{array}{lll}
U={\dis p v \, \frac{(1-q)u+q-p+(p-1)v}{q(1-p) u+p(q-1)v+(p-q)uv}},
                                                  & V={\dis q u \, \frac{(1-q)u+q-p+(p-1)v}{q(1-p) u+p(q-1)v+(p-q)uv}}
                                                                                                       & (F_{I})         \\ [3mm]
U={\dis \frac{v}{p} \,  \frac{pu-qv+q-p}{u-v}},     & V={\dis \frac{u}{q} \,  \frac{pu-qv+q-p}{u-v}}        & (F_{II})        \\ [3mm]
U={\dis \frac{v}{p} \,  \frac{pu-qv}{u-v}},     & V={\dis \frac{u}{q} \,  \frac{pu-qv}{u-v}}                 & (F_{III})       \\ [3mm]
U={\dis v\left(1+\frac{p-q}{u-v}\right)},         & V={\dis u \left(1+\frac{p-q}{u-v}\right)},         & (F_{IV})        \\ [3mm]
U={\dis v + \frac{p-q}{u-v}},                     & V={\dis u + \frac{p-q}{u-v}},                      & (F_V)
\end{array}
\end{displaymath}
as well as the companions of  H-list (cH-list) of quadrirational  Yang-Baxter maps  presented in \cite{PSTV}
\begin{displaymath}
\label{cH}
\begin{array}{lll}
U={\dis \frac{q(1-p) u+p(q-1)v+(p-q)uv}{v [(1-q)u+q-p+(p-1)v]}},&
V={\dis \frac{q(1-p) u+p(q-1)v+(p-q)uv}{u [(1-q)u+q-p+(p-1)v]}},& (cH_I)\\ [3mm]
U={\dis \frac{1-v}{p} \, \frac{pu-qv}{ u-v}},&V={\dis \frac{1-u}{q} \, \frac{pu-qv }{u-v }}&(cH_{II})\\ [3mm]
U={\dis -\frac{v}{p} \, \frac{pu-qv}{u-v}},&V={\dis -\frac{u}{q} \, \frac{pu-qv}{u-v}}& (cH_{III}^A)\\ [3mm]
U={\dis-\frac{1}{v} \, \frac{u-v}{pu-qv}},& V={\dis -\frac{1}{u} \, \frac{u-v}{pu-qv}}&(cH_{III}^B)\\ [3mm]

U={\dis -v+\frac{p-q}{u-v}},& V={\dis -u+ \frac{p-q}{u-v}}& (cH_V)
\end{array}
\end{displaymath}
where it is worth mentioning that the companion maps  are not  Yang-Baxter maps.
The two lists of maps are related by $(M\ddot{o}b)^4$-action.
In particular,
the change of variables
\begin{equation}
(U,V,u,v) \mapsto (-U,-V,u,v),
\end{equation} transforms map $F_V$  into map $cH_V$ with parameters  $(p,q)$  replaced with $(-p,-q)$,  as well as it transforms map $F_{III}$ into  map $cH_{III}^A.$
The map
\begin{equation}
(U,V,u,v) \mapsto (-1/(pU),-1/(qV),u,v),
\end{equation} transforms map $F_{III}$  into map $cH_{III}^B.$
The map
\begin{equation}
(U,V,u,v) \mapsto  (p/U,q/V,u,v),
\end{equation}
 transforms
the  $F_I$ map into  the $cH_I$ map.
Finally
the map 
\begin{equation}
 (U,V,u,v) \mapsto,  (1-U,1-V,u,v)
 \end{equation}
transforms  map $F_{II}$  into map  $cH_{II}.$

Three comments are in order. First, although both lists yield, as we shall see in the next section, essentially the same (i.e. point equivalent)   structures,
we decided to present results concerning both lists for the equivalence is not always obvious.
Second, the advantage of using  cH-list  rather than H list lies in the fact that  cH-list can be extended to multidimensions (see Section \ref{Section5}) in full analogy to F-list whereas H-list cannot be extended this way. Third, all the maps presented above are involutions.


\section{Invariants or alternating invariants with separated variables} \label{Section3}

Following the procedure described in \cite{KaNie}  for each of the maps  from Subsection \ref{ss22} we present the  exhaustive list of invariants and alternating invariants with separated variables.  For each map the number of linearly independent invariants and alternating invariants in separated form is three. In the next section we will present an alternative derivation of the invariants with separated variables, this derivation  leads to a first order differential equation and allows us for another characterization of the map. The exhaustive list of invariants and alternating invariants for the maps of subsection \ref{subsec2.1} was derived in \cite{KaNie:2011}.

The results are included in Table \ref{inttable}.
We underline that the invariants (or alternating invariants) can be combined e.g. for map $F_{III}$ function
$I(u,v)=a (\ln \sqrt{p} u- \ln \sqrt{q} v)+b(pu-qv)+c(\frac{1}{u}-\frac{1}{v})+d$ where $a,b,c,d$ are some constants is alternating integral with separated variables as well.
\begin{table}[h!]
\caption{Maps and their invariants (grey background) or alternating invariants (white background) with separated variables. To avoid the appearance of logarithms in formulas of the (alternating) invariants, some of the (alternating) invariants are presented in exponentiated form (the separation of variables is multiplicative)}
\label{inttable}
\begin{tabular}{l|lll}
\toprule
Map &  &  &\\ [3mm]\hline
$F_{I}$
      & $\sqrt{\frac{q}{p}}\frac{U}{V}=\sqrt{\frac{p}{q}} \frac{ v}{u}$
      & $ \sqrt{\frac{q-1}{p-1}}\frac{U-1}{V-1}=\sqrt{\frac{p-1}{q-1}} \frac{v-1}{u-1}$
      & $\sqrt{\frac{q(q-1)}{p(p-1)}}\frac{U-p}{V-q}=\sqrt{\frac{p(p-1)}{q(q-1)}} \frac{v-q}{u-p}$\\ [3mm]
$F_{II}$
      & $\sqrt{\frac{p}{q}} \frac{U}{V}=\sqrt{\frac{q}{p}} \frac{v}{u}$
      & $\sqrt{\frac{p}{q}} \frac{U-1}{V-1}= \sqrt{\frac{q}{p}}\frac{v-1}{u-1}$
      & {\scriptsize $pU-\frac{1}{2}p-qV+\frac{1}{2}q=-(p u-\frac{1}{2}p-q v+\frac{1}{2}q) $}\\ [3mm]
$F_{III}$
      &  $\sqrt{\frac{p}{q}}\frac{U}{V}=\sqrt{\frac{q}{p}}\frac{v}{u}$
      & {\scriptsize $\frac{1}{U}-\frac{1}{V}=-(\frac{1}{u}-\frac{1}{v})$ }
      &  $pU-qV=-(pu-qv)$\\ [3mm]
$F_{IV}$
      &  $\frac{U}{V}=\frac{v}{u}$
      &  {\scriptsize $U+\frac{1}{2}p-V-\frac{1}{2} q=-(u+\frac{1}{2}p-v-\frac{1}{2} q)$ }
      &  \parbox[t]{6cm}{\scriptsize $U^2+2pU+\frac{p^2}{2}-V^2-2qV-\frac{q^2}{2}=$  \\ \hbox{~~~ ~~~ ~~~ ~~~ ~~~} $ -(u^2+2pu+\frac{p^2}{2}-v^2-2qv-\frac{q^2}{2})$}\\ [3mm]
$F_{V}$
      & \parbox[t]{1.2cm}{\scriptsize  $U-V=$ \\ \hbox{~} $-(u-v)$}
      &  {\scriptsize $U^2+p-V^2-q=-(u^2+p-v^2-q)$}
      &  {\scriptsize $U^3+3pU-V^3-3qV=-(u^3+3pu-v^3-3qv)$}\\ [3mm]\hline
$cH_{I}$
      &\cellcolor[gray]{0.85}  $\frac{U}{V}=\frac{u}{v}$
      &\cellcolor[gray]{0.85}  $ \frac{(U-1)}{(V-1)}  \frac{(V-q)}{(U-p)}= \frac{(u-1)}{(v-1)} \frac{(v-q)}{(u-p)} $
      & $\frac{(q-1)}{(p-1)}\frac{V}{U} \frac{(U-1)}{(V-1)}  \frac{(U-p)}{(V-q)}= \frac{(p-1)}{(q-1)}\frac{u}{v}\frac{(v-1)}{(u-1)} \frac{(v-q)}{(u-p)} $\\ [3mm]
$cH_{II}$
      & \cellcolor[gray]{0.85}  $\frac{U(V-1)}{V(U-1)}=\frac{u(v-1)}{v(u-1)}$
      & \cellcolor[gray]{0.85}  $pU-qV=pu-qv$
      & $\frac{pU(U-1)}{qV(V-1)}=\frac{qv(v-1)}{pu(u-1)} $\\ [3mm]
$cH_{III}^A$
      &  $\sqrt{\frac{p}{q}}\frac{U}{V}=\sqrt{\frac{q}{p}}\frac{v}{u}$
      &  \cellcolor[gray]{0.85} {\scriptsize $pU-qV=pu-qv$}
      &  \cellcolor[gray]{0.85} $\frac{1}{U}-\frac{1}{V}=\frac{1}{u}-\frac{1}{v}$\\ [3mm]
$cH_{III}^B$
          & \cellcolor[gray]{0.85} $\frac{U}{V}=\frac{u}{v}$
         &\cellcolor[gray]{0.85} $pU+\frac{1}{U}-qV-\frac{1}{V} = pu+\frac{1}{u}-qv-\frac{1}{v}$
          &  $pU-\frac{1}{U}-qV+\frac{1}{V} =-\left( pu-\frac{1}{u}-qv+\frac{1}{v}\right)$
 \\ [3mm]
$cH_{V}$
      & \cellcolor[gray]{0.85} \parbox[t]{1.2cm}{\scriptsize  $U-V=$ \\ \hbox{~~~~~} $u-v$ }
      &  {\scriptsize $U^2-p-V^2+q=-(u^2-p-v^2+q)$}
      &  \cellcolor[gray]{0.85} {\scriptsize $U^3-3pU-V^3+3qV=u^3-3pu-v^3+3qv$}\\ [3mm]
\bottomrule
\end{tabular}
\end{table}

\newpage

\section{Finding (alternating) invariants with separated variables} \label{Section4}

All the maps in question admit a  method of finding  all of  their (alternating) invariants with separated variables
which is simpler than the general method described in \cite{KaNie}.
The method leads to the differential equation
\begin{equation}
\label{diffE}
\frac{d F(u)}{du}= \frac{b_0+b_1 u+ b_2 u^2}{c_0+c_1 u+ c_2 u^2+c_3 u^3 +c_4 u^4}
\end{equation}
covariant under a homographic transformation $u\mapsto h(u)=\frac{\alpha u+\beta}{\gamma u+\delta}$.

Here we consider a more general problem. We search for functions $F, G, f$ and $g$ such that
\begin{equation}
\label{FGfg}
F(U(u,v))+G(V(u,v))= f(u)+g(v).
\end{equation}
 We take any of the maps of Section \ref{Section2}. After the differentiation of (\ref{FGfg}) with respect to $u$ and $v,$ the right hand side of (\ref{FGfg}) vanishes   and using the inverse of the  map under consideration,  we obtain a relation of the form
$$
Q(F''(U),F'(U),G''(V),G'(V),U,V;p,q)=0.
$$
Differentiating the relation above  with respect to $U$ and $V$ yields
\be \label{seper}
m(U)F'''(U)+2m'(U)F''(U)+m''(U)F'(U)=-n(V)G'''(V)-2n'(V)G''(V)-n''(V)G'(V),
\ee
 where $m(U)$  and $n(V)$ are polynomials depending on the  map under consideration and are given explicitly in Table \ref{uvtable}. As one can notice the variables in equation (\ref{seper}) separate and hence both sides of (\ref{seper}) should be constant, namely
 $$
 m(U)F'''(U)+2m'(U)F''(U)+m''(U)F'(U)=c, \quad -n(V)G'''(V)-2n'(V)G''(V)-n''(V)G'(V)=c.
 $$
The differential equations above can be integrated twice to give the following equations
\be \label{riccati}
m(U)F'(U)=\frac{c}{2}U^2+d U+e,\quad n(V)G'(V)=-\frac{c}{2}V^2+k V+l
\ee

\begin{table}[h!]
\caption{The polynomials $m(u)$ and $n(v)$ of equation (\ref{riccati}) associated with the F and the cH-list of maps}
\label{uvtable}
\begin{tabular}{l|ll}
\toprule
Map & m(u) & n(v) \\ [3mm]\hline
$F_{I}$   & $u( u-1) (p-u)$      & $ v( v-1) (q-v)$      \\ [3mm]
$F_{II}$
      & $u(u-1)/p$
      & $v (v-1)/q$
      \\ [3mm]
$F_{III}$
      &  $u^2/p$
      & $v^2/q$\\ [3mm]
$F_{IV}$
      &  $u$
      &  $v$\\ [3mm]
$F_{V}$
      & $1$
      &  $1$\\ [3mm]\hline
$cH_{I}$
      &$u(u-1)(p-u)$ &$v( v-1) (q-v)$\\ [3mm]
$cH_{II}$
      &$u(u-1)/p$ &$v(v-1)/q$\\ [3mm]
$cH_{III}^A$
      & $u^2/p$ &$v^2/q$\\ [3mm]
$cH_{III}^B$
          & $u^2/p$ &$v^2/q$ \\ [3mm]
$cH_{V}$
      & 1&  1    \\ [3mm]
\bottomrule
\end{tabular}
\end{table}
It is not difficult to see that equations \mref{riccati} can be obtained from equation \mref{diffE} by means of
homographic transformation by sending one root to infinity.
The cases when the polynomial   $c_0+c_1 u+ c_2 u^2+c_3 u^3 +c_4 u^4$ in \mref{diffE}  has four roots, three roots, two double roots,  one triple root and one single root, one quadruple root,
reflect the classification of the maps leading to  maps with the Roman subscript $I$, $II$, $III$, $IV$, $V$ respectively.

\newpage
\section{Difference equations} \label{Section5}
\label{Systems}
The next  step is to reinterpret the map $\Phi:(u,v)\mapsto (U,V)$,  as a difference system on the $\mathbb{Z}^2$ lattice. This is achieved by assigning the map  on every elementary square of the $\mathbb{Z}^2$ lattice, where the variables $u,v$ are now considered as functions (in independent variables $(m_1,n_1)\in{\mathbb Z}^2$ and $(m_2,n_2)\in{\mathbb Z}^2$ respectively), assigned to two   perpendicular edges with a common vertex of an elementary square of the $\mathbb{Z}^2$ lattice and $U,V$ to the  corresponding opposite  edges, so $u\equiv u_{m_1,n_1}, v\equiv v_{m_2,n_2},$ $U\equiv u_{m_1,n_1+1}$ and $V\equiv u_{m_2+1,n_2}$ (see Figure \ref{notation0}). More precisely, as we have already mention in introductory Section \ref{Section0},  we consider $u$ as a function  defined  on the set of horizontal edges
i.e. pairs of vertices $(m_1,n_1):=\{(m,n),(m+1,n)\}, (m,n) \in {\mathbb Z}^2 $
and respectively $v$ as a function  defined  on set of vertical edges i.e. pairs of vertices $(m_2,n_2):=\{(m,n),(m,n+1)\}, (m,n)  \in {\mathbb Z}^2$, see Figure \ref{notation1} $(a)$. Moreover, we introduce a concise notation, first, by considering  functions $u^i, p^i, i=1,2$ such that
 $u^1_{m_1,n_1}:=u_{m_1,n_1},$ $u^2_{m_2,n_2}:=v_{m_2,n_2},$ $p^1_{m_1}:=p_{m_1}$ and $p^2_{n_2}:=q_{n_2}$, second,  we omit the dependency of the functions $u^i, p^i, i=1,2$ on the independent variables, third, we denote the shifts in the corresponding variables as subscripts i.e. $u^i_1=u^i_{m_i+1,n_i}, u^i_2=u^i_{m_i,n_i+1}, u^i_{11}=u^i_{m_i+2,n_i},$ etc.
 Therefore, with  this concise notation we enumerate functions with superscripts and the shifts with subscripts, see Figure \ref{notation1} $(b)$.
For the cH-list we will use the letter $s$ instead of $u$ to denote the dependent variable.

\begin{figure}[h]
\begin{minipage}[h]{0.5\textwidth}
\begin{tikzpicture}[scale=2, every node/.style={transform shape}];
\draw[ultra thick] (-0.5,0)--(-0.2,0)  (0.5,0) node [scale=0.4, above ] { $ u_{m_1,n_1}$}     (0.5,0) node [scale=0.4, below ] { $ p_{m_1}$};
\draw[ultra thick] (0.2,0)--(0.8,0);
\draw[ultra thick] (1.2,0)--(1.8,0)    (1.5,0) node [scale=0.4, above ] { $ u_{m_1+1,n_1}$} (1.5,0) node [scale=0.4, below ] { $ p_{m_1+1}$};
\draw[ultra thick] (2.2,0)--(2.5,0);
\draw[ultra thick] (-0.5,1)--(-0.2,1)  (0.5,1) node [scale=0.4, above ] { $ u_{m_1,n_1+1}$} (0.5,1) node [scale=0.4, below ] { $ p_{m_1}$};
\draw[ultra thick] (0.2,1)--(0.8,1);
\draw[ultra thick] (1.2,1)--(1.8,1)    (1.5,1) node [scale=0.4, above ] { $ u_{m_1+1,n_1+1}$}  (1.5,1) node [scale=0.4, below ] { $ p_{m_1+1}$};;
\draw[ultra thick] (2.2,1)--(2.5,1);
\draw[ultra thick] (-0.5,2)--(-0.2,2);
\draw[ultra thick] (0.2,2)--(0.8,2);
\draw[ultra thick] (1.2,2)--(1.8,2);
\draw[ultra thick] (2.2,2)--(2.5,2);
\draw  (0,0) node [scale=0.3, fill=white ] { $ \psi_{m,n}$};
\draw    (1,0) node [scale=0.3, fill=white ] { $\psi_{m+1,n}$};
\draw    (2,0) node [scale=0.3, fill=white]  { $\psi_{m+2,n}$};
\draw    (0,1) node [scale=0.3, fill=white ] {$\psi_{m,n+1}$};
\draw    (0,2) node [scale=0.3, fill=white ] {  $\psi_{m,n+2}$};
\draw    (1,1) node [scale=0.3, fill=white ] { $\psi_{m+1,n+1}$};

\draw[ultra thick] (0,-0.5)--(0,-0.2)   (-0.08,0.7) node [scale=0.4, left, rotate=90 ] {$v_{m_2,n_2}$}  (0.92,0.75) node [scale=0.4, left, rotate=90 ] {$v_{m_2+1,n_2}$} (0.08,0.3) node [scale=0.4, right, rotate=90 ] {$q_{n_2}$} (1.08,0.3) node [scale=0.4, right, rotate=90 ] {$q_{n_2}$} (0.08,1.25) node [scale=0.4, right, rotate=90 ] {$q_{n_2+1}$} (1.08,1.25) node [scale=0.4, right, rotate=90 ] {$q_{n_2+1}$};
\draw[ultra thick] (0,0.1)--(0,0.8)  (-0.08,1.8) node [scale=0.4, left, rotate=90 ] {$v_{m_2,n_2+1}$}  (0.92,1.9) node [scale=0.4, left, rotate=90 ] {$v_{m_2+1,n_2+1}$} ;
\draw[ultra thick] (0,1.2)--(0,1.8);
\draw[ultra thick] (0,2.2)--(0,2.5);
\draw[ultra thick] (1,-0.5)--(1,-0.2);
\draw[ultra thick] (1,0.2)--(1,0.8);
\draw[ultra thick] (1,1.2)--(1,1.8);
\draw[ultra thick] (1,2.2)--(1,2.5);
\draw[ultra thick] (2,-0.5)--(2,-0.2);
\draw[ultra thick] (2,0.2)--(2,0.8);
\draw[ultra thick] (2,1.2)--(2,1.8);
\draw[ultra thick] (2,2.2)--(2,2.5);
%
\end{tikzpicture}
\captionsetup{font=footnotesize}
\captionof*{figure}{(a)}
\end{minipage}
\begin{minipage}[h]{0.5\textwidth}
\begin{tikzpicture}[scale=2, every node/.style={transform shape}];
\draw  (0,0) node [scale=0.3 ] { $ \psi$};
\draw    (1,0) node [scale=0.3 ] { $\psi_{1}$};
\draw    (2,0) node [scale=0.3]  { $\psi_{11}$};
\draw    (0,1) node [scale=0.3 ] {$\psi_{2}$};
\draw    (0,2) node [scale=0.3 ] {  $\psi_{22}$};
\draw    (1,1) node [scale=0.3 ] { $\psi_{12}$};
\draw[ultra thick] (-0.5,0)--(-0.2,0)  (0.5,0) node [scale=0.5, above ] { $ u^1$} node [scale=0.5, below ] { $ p^1$};
\draw[ultra thick] (0.2,0)--(0.8,0);
\draw[ultra thick] (1.2,0)--(1.8,0)    (1.5,0) node [scale=0.5, above ] { $ u^1_1$} node [scale=0.5, below ] { $ p^1_1$};
\draw[ultra thick] (2.2,0)--(2.5,0);
\draw[ultra thick] (-0.5,1)--(-0.2,1)  (0.5,1) node [scale=0.5, above ] { $ u^1_{2}$} node [scale=0.5, below ] { $ p^1$} ;
\draw[ultra thick] (0.2,1)--(0.8,1);
\draw[ultra thick] (1.2,1)--(1.8,1)    (1.5,1) node [scale=0.5, above ] { $ u^1_{12}$} node [scale=0.5, below ] { $ p^1_1$};
\draw[ultra thick] (2.2,1)--(2.5,1);
\draw[ultra thick] (-0.5,2)--(-0.2,2);
\draw[ultra thick] (0.2,2)--(0.8,2);
\draw[ultra thick] (1.2,2)--(1.8,2);
\draw[ultra thick] (2.2,2)--(2.5,2);
\draw[ultra thick] (0,-0.5)--(0,-0.2)   (-0.2,0.6) node [scale=0.5, left, rotate=90 ] {$u^2$} (0.2,0.33) node [scale=0.5, right, rotate=90 ] {$p^2$}  (0.8,0.6) node [scale=0.5, left, rotate=90 ] {$u^2_1$} (1.2,0.33) node [scale=0.5, right, rotate=90 ] {$p^2$} ;
\draw[ultra thick] (0,0.2)--(0,0.8)  (-0.2,1.6) node [scale=0.5, left, rotate=90 ] {$u^2_2$} (0.2,1.3) node [scale=0.5, right, rotate=90 ] {$p^2_2$}  (0.8,1.7) node [scale=0.5, left, rotate=90 ] {$u^2_{12}$} (1.2,1.4) node [scale=0.5, right, rotate=90 ] {$p^2_2$};
\draw[ultra thick] (0,1.2)--(0,1.8);
\draw[ultra thick] (0,2.2)--(0,2.5);
\draw[ultra thick] (1,-0.5)--(1,-0.2);
\draw[ultra thick] (1,0.2)--(1,0.8);
\draw[ultra thick] (1,1.2)--(1,1.8);
\draw[ultra thick] (1,2.2)--(1,2.5);
\draw[ultra thick] (2,-0.5)--(2,-0.2);
\draw[ultra thick] (2,0.2)--(2,0.8);
\draw[ultra thick] (2,1.2)--(2,1.8);
\draw[ultra thick] (2,2.2)--(2,2.5);
\end{tikzpicture}
\captionsetup{font=footnotesize}
\captionof*{figure}{(b)}
\end{minipage}
\caption{Fields on  the edges of the ${\mathbb Z}^2$ lattice. \newline (a): Standard notation. (b): concise notation used in the paper}\label{notation1}
\end{figure}

%
%
E.g. for the map $F_V$ from the list \mref{F} we get \cite{Tasos}
\begin{equation}
\label{uu}
 u^i_j={\dis u^j + \frac{p^i-p^j}{u^i-u^j}},  \quad i,j=1,2, \quad i\neq j.
\end{equation}

          All the difference systems obtained this way from maps of subsection \ref{ss22}, can  be extended to multi-dimensions,  see \cite{ABS,KaNie,KaNie:2011}. Namely, allowing in \mref{uu}
$i,j=1, \ldots, n$
we observe that the following compatibility conditions hold
\begin{equation}  \label{compa}
u^i_{jk}=u^i_{kj}, \quad s^i_{jk}=s^i_{kj} \quad i,j,k=1, \ldots, n , \qquad i \neq j \neq k \neq i.
\end{equation}

Invariants which were presented in  Section \ref{Section3}, turn into equations that guarantee the existence of potentials. By a potential we understand a function $\psi$ such that
\begin{equation}\label{ff}
\psi_i \pm \psi=f(u^i,p^i), \quad i=1, \ldots, n.
\end{equation}
Therefore, we obtain a list of the difference systems 
as well as their four-parameter families of  scalar potentials. The results are collected in Tables \ref{IsF} and \ref{IsH}, where in order to stress the fact that we deal with multi-dimensional systems of difference equations and not maps, we introduce the $I$-list as the multidimensional extension of the $F$-list and the $cI$-list as the multidimensional extension of the $cH-$list.
\begin{table}[!h]
\caption{ The Table includes  equations associated with the $F$ list of maps and expressions $\psi_{i}+\psi$ which define families of potentials $\psi$ by discrete quadratures}
\label{IsF}
\begin{tabular}{lll}
\toprule
Label &  Difference equation & Expression $\psi_{i}+\psi$ \\
\hline
$I_{I}$ & $ u^i_j = p^iu^j{\dis \frac{(1-{p^j}) u^i - (1-{p^i}) u^j-{p^i}+{p^j}}{p^j (1-{p^i}) u^i - p^i (1-{p^j}) u^j+({p^i}-{p^j})u^iu^j}}$
&
$a \ln \frac{u^i}{\sqrt p^i}+b\ln \frac{u^i-1}{\sqrt{{p^i}-1}} +c\ln \frac{u^i-p^i}{\sqrt{p^i({p^i}-1)}}+d$
\\
$I_{II}$ & $u^i_j = \frac{u^j}{p^i} {\dis\frac{p^i u^i - p^j u^j-{p^i}+{p^j}}{u^i - u^j}}$
&
$a \ln {\sqrt p^i}u^i +b \ln {\sqrt p^i}(u^i-1)$+$c(2 p^i u^i- {p^{i}})+d$
\\
$I_{III}$ & $ u^i_j = \frac{u^j}{p^i} {\dis\frac{p^i u^i - p^j u^j}{u^i - u^j}}$
&
$a \ln {\sqrt p^i}u^i+ \frac{b}{u^i}+c p^iu^i+d$
\\
$I_{IV}$ & $u^i_j = u^j {\dis (1+\frac{p^i  - p^j }{u^i - u^j})}$
&
$a \ln u^i+ b(2  u^i + p^i) +c [({u^i} + p^i)^2-\frac{{p^i}^2}{2}]+d$
\\
$I_{V}$  & $u^i_j = u^j +{\dis\frac{p^i  - p^j }{u^i - u^j}}$
&
$a  u^i +b ({u^i}^2+p^i)+c({u^i}^3+3p^i u^i)+d$
\\
\bottomrule
\end{tabular}
\caption{The Table includes  equations associated with the $cH$-list of maps and expressions $\psi_{i}-\psi$ which define families of potentials $\psi$ by discrete quadratures.
$\sigma:= (-1)^{m_1+\ldots +m_n}$}
\label{IsH}
\begin{tabular}{lll}
\toprule
label & Difference equation & Expression  $\psi_i-\psi =$\\
\hline
$cI_{I}$ & $s^{i}_j={\dis \frac{p^j(1-p^i)s^i+p^i (p^j-1)s^j+(p^i-p^j)s^is^j}{s^j(s^i-s^j-p^js^i+p^is^j+p^j-p^i)}}$ & $ a\ln s^i+b\ln \left(\frac{s^i-1}{s^i-p^i}\right) -c \sigma\ln\frac{(s^i-1)(s^i-p^i)}{s^i(p^i-1)}+d$\\
$cI_{II}$ & $s^{i}_j={\dis \frac{1-s^j}{p^i}\frac{p^is^i-p^js^j}{s^i-s^j}}$ & $ a\ln \frac{s^i}{s^i-1}+bp^is^i-c\sigma\ln p^i s^i (s^i-1)+d $\\
$cI_{III}^A$ & $ s^{i}_j= {\dis -\frac{s^j}{p^i} \frac{p^is^i-p^js^j}{s^i-s^j}} $ & $ -a \sigma \ln {\sqrt p^i}s^i+b p^i s^i+c \frac{1}{s^i}+d $ \\
$cI_{III}^B$ & $ s^{i}_j= {\dis -\frac{1}{s^j} \frac{s^i-s^j}{p^is^i-p^js^j}} $  &   $ a \ln s^i+b\left(p^is^i+\frac{1}{s^i}\right)
-c\sigma\left(p^is^i-\frac{1}{s^i}\right)+d$  \\
$cI_{V}$ & $ s^{i}_j= -s^j+{\dis\frac{p^i-p^j}{s^i-s^j} }  $ & $   as^i-b\sigma\left({s^i}^2-p^i\right)+c \left({s^i}^3-3 p^is^i\right)+d$ \\
\bottomrule
\end{tabular}
\end{table}

\newpage

The equations with the same Roman index in Tables \ref{IsF} and \ref{IsH} are point equivalent  and the transformations between them are given in  Table
\ref{IvcI}.
\begin{table}[!h]
\caption{The point transformations between difference equations from Tables \ref{IsF} and \ref{IsH}}
\label{IvcI}
\begin{tabular}{lll}
\toprule
Equation in $u^i$ & A point transformation & Equation in $s^i$ \\
\hline
$I_I$ & $u^i={p^i}^{\frac{1-\sigma}{2}}{s^i}^{\sigma}$ & $cI_I$ \\
$I_{II}$ & $ u^i= {\frac{1-\sigma}{2}}+\sigma s^i$ & $cI_{II}$ \\
$I_{III}$ & $ u^i= \sigma s^i $ & $cI^A_{III}$ \\
$I_{III}$ & $ u^i= \sigma {p^i}^{\frac{-1+\sigma}{2}}{s^i}^{\sigma}$ & $cI^B_{III} $ \\
$I_{V}$ & $ u^i=\sigma s^i$ & $cI_{V}\,\,\, (-p^i)$ \\
\bottomrule
\end{tabular}
\end{table}

Finally from the difference equations given in Tables \ref{IsF} and \ref{IsH} and from the expressions  $\psi_i + \psi=f(u^i;p^i)$ ($\psi_i - \psi=f(s^i;p^i)$) one can derive expressions $\psi_{ij} - \psi=g(u^i,u^j;p^i,p^j)$   ($\psi_{ij} - \psi=g(s^i,s^j;p^i,p^j)$) which in turn allows us (see  Section \ref{Idolons}) to rewrite the difference equations in terms of their potentials.  Tables
\ref{isF} and \ref{isH} include the expressions $\psi_{ij} - \psi$ in terms of the edge variables.
\begin{table}[!h]
\caption{Expressions $\psi_{ij} - \psi=g(u^i,u^j;p^i,p^j)$ for the $I$ systems  from Table \ref{IsF}}
\label{isF}
\begin{tabular}{ll}
\toprule
Equation & Expression $\psi_{ij}-\psi$\\
\hline
$I_{I}$
&
$ a\ln \sqrt{p^ip^j} \frac{(1-p^j)u^i+p^j-p^i-(1-p^i)u^j}{p^j(1-p^i)u^i-p^i(1-p^j)u^j+(p^i-p^j)u^iu^j}+
b\ln \sqrt{(p^j-1)(p^i-1)} \frac{p^iu^j-p^ju^i}{p^j(1-p^i)u^i-p^i(1-p^j)u^j+(p^i-p^j)u^iu^j}+$\\
& $+c\ln \sqrt{p^ip^j(p^i-i)(p^j-1)} \frac{u^j-u^i}{p^j(1-p^i)u^i-p^i(1-p^j)u^j+(p^i-p^j)u^iu^j}$
\\ [2mm]
$I_{II}$
&
$
a \ln \frac{1}{\sqrt{p^ip^j}} \frac{p^iu^i-p^ju^j-{p^i}+{p^j}}{ u^i- u^j}+
b \ln \frac{1}{\sqrt{p^ip^j}} \frac{p^iu^i-p^ju^j}{u^i-u^j}+c (p^i-p^j)\frac{2u^iu^j-u^i-u^j}{u^i-u^j}
$
 \\ [2mm]
$I_{III}$
&
$a \ln \frac{1}{\sqrt{p^ip^j}}\frac{p^i u^i- p^j u^j}{u^i-u^j} +({p^j}-{p^i}) \left(b \frac{1}{p^iu^i-p^ju^j}- c \frac{u^iu^j}{u^i-u^j}  \right)$
\\ [2mm]
$I_{IV}$
&
$ a \ln \left(1+ \frac{p^i-p^j}{u^i-u^j} \right)+ b\frac{p^i  - p^j }{u^i - u^j} (u^i + u^j)
+c  \frac{p^i-p^j}{u^i-u^j}[2u^i u^j+\frac{p^i-p^j}{u^i-u^j}u^i u^j +\frac{1}{2}(p^i+p^j)(u^i + u^j) ]$
 \\ [2mm]
$I_{V}$
&
$(p^i-p^j)\left[\frac{\psi_i - \psi_j}{(u^i - u^j)^2} +c\left(\frac{(p^i-p^j)^2}{(u^i - u^j)^3}-u^i+u^j\right)  \right]$
\\
\bottomrule
\end{tabular}
\caption{Expressions $\psi_{ij} - \psi=g(s^i,s^j;p^i,p^j)$ for the  $cI$ systems  from Table \ref{IsH} 
}
\label{isH}
\begin{tabular}{ll}
\toprule
Equation & Expression $\psi_{ij}-\psi$\\
\hline
$cI_{I}$ & $  a \ln \frac{p^j(1-p^i)s^i-p^i(1-p^j)s^j+(p^i-p^j)s^is^j}{s^i-s^j+p^j(1-s^i)-p^i(1-s^j)}+b \ln \frac{s^i-s^j}{p^js^i-p^is^j}+$ \\ [2mm]
& $+ c \sigma
\ln \frac{(p^i-1)(p^j-1)(s^j-s^i)(p^is^j-p^js^i)}{\left[s^i-s^j+p^j(1-s^i)-p^i(1-s^j)\right]
\left[p^j(1-p^i)s^i-p^i(1-p^j)s^j+(p^i-p^j)s^is^j\right]}$ \\ [2mm]
$cI_{II}$ & $a\ln \frac{p^is^i-p^js^j}{p^j(1-s^j)-p^i(1-s^i)} +b\frac{p^is^i(1-s^j)-p^js^j(1-s^i)}{s^i-s^j}+ {c}\sigma
\ln \frac{(p^is^i-p^js^j)\left(p^j(1-s^j)-p^i(1-s^i)\right)}{p^ip^j(s^i-s^j)^2} $ \\ [2mm]
$cI_{III}^A$ & $  a \sigma \ln \frac{1}{\sqrt{p^ip^j}}\frac{p^is^i-p^js^j}{s^i-s^j} +b \frac{({p^j}-{p^i})s^is^j}{s^i-s^j} +c \frac{{p^i}-{p^j}}{p^is^i-p^js^j} $
\\ [2mm]
$cI_{III}^B$ & $ a\ln \frac{s^j-s^i}{p^is^i-p^js^j} +(b+c\sigma)\frac{p^i-p^j}{p^is^i-p^js^j}  -(b-c\sigma)(p^i-p^j)\frac{s^is^j}{s^i-s^j}$
\\ [2mm]
$cI_{V}$ & $(p^i-p^j)\left[\frac{\psi_i - \psi_j}{(s^i - s^j)^2} +c\left(\frac{(p^i-p^j)^2}{(s^i - s^j)^3}-s^i+s^j\right)  \right]$
\\
\bottomrule
\end{tabular}
\end{table}

\section{Idolons} \label{Section6}
\label{Idolons}

We would like to emphasize the superior role of the difference systems defined on edges (bond systems)  described so far,  by referring to  (adopting Plato's terminology) any vertex system that arises from rewriting the bond system in terms of its particular potential as idolon  (i.e. we always treat the idolons as coming from the superior bond systems).  By not  treating the vertex models (idolons) alone but remembering that they arise from the corresponding bond systems, the issue of non-single valued  evolution (if it appears) can be resolved.

\subsection{General situation}
 Taking system $I_{V}$ as an example, we illustrate the advantages of non treating the correspondences governing potentials in isolation from the difference system they come from.
The family of potentials $\psi$ of $I_{V}$
obeys
\begin{equation}
\label{evo}
\psi_{ij}-\psi=
(p^i-p^j)\left[\frac{\psi_i - \psi_j}{(u^i - u^j)^2} +c\left(\frac{(p^i-p^j)^2}{(u^i - u^j)^3}-u^i+u^j\right) \right],
\end{equation}
where functions $u^i$ are given implicitly by
\begin{equation}
\label{impl}
\psi_{i}+\psi=a  u^i +b ({u^i}^2+p^i)+c({u^i}^3+3p^i u^i)+d.
\end{equation}
It is  the implicit relation \mref{impl} what gives the  multivaluedness of the system.
We come with the proposition not to consider \mref{evo} and \mref{impl} alone but supplement them with the
equations
\begin{equation}
\label{supl}
u^i_j = u^j +{\dis\frac{p^i  - p^j }{u^i - u^j}}
\end{equation}
i.e. we consider only such solutions of \mref{evo} and \mref{impl} that admit existence of functions $u^i$ that obeys \mref{supl}.
It makes the system single-valued provided that a well-possed initial value problem is given.
One of such well-possed initial value problem is  to prescribe the function $\psi$ at one point (say ${\mathcal P}=(m^1_0,\ldots , m^n_0)$) and the functions $u^i$ on edges
that belong to straight lines passing through ${\mathcal P}$ and with all coordinates but one fixed (so function $u^i$ is prescribed on $n$ lines, where the $i-$th line consists of set of edges\\
$\left\{ ((m^1_0,\ldots , m^i_0+k, \ldots ,m^n_0),(m^1_0,\ldots , m^i_0+k+1, \ldots ,m^n_0))\, | \, k\in {\mathbb Z}\right\},$ see Figure \ref{init1} where $n=2$).
A second example of such initial value problem is  to prescribe a function $\psi$ at one point ${\mathcal P}$ and the functions $u^i$ on edges, so that both edges and vertices  belong to the  strip (staircase) $\{(m^1, \ldots ,m^n)\in  {\mathbb Z}^n \, | \,m^1_0 +\ldots+m_0^n \leq m^1+\ldots+m^n< m^1_0 +\ldots+m_0^n +n\},$
see Figure \ref{init2} where $n=2$.

\begin{figure}[h]
\begin{minipage}{0.2\linewidth} \label{fig1}
\begin{texdraw}
\setunitscale 0.3
\move(0 0) \lvec(1 0) \move(1.1 0) \lvec(2 0) \move(2.1 0) \lvec(3 0) \fcir f:0  r:0.08  \move(3.1 0) \lvec(4 0) \move(4.1 0) \lvec(5 0) \move(5.1 0)
\lvec(6 0)

\move(3 0) \lvec(3 1) \move(3 1.1) \lvec(3 2) \move(3 2.1) \lvec(3 3)
\move(3 0) \lvec(3 -1) \move(3 -1.1) \lvec(3 -2) \move(3 -2.1) \lvec(3 -3)

\end{texdraw}
\caption{ Intersecting lines initial value problem} \label{init1}
\end{minipage}
\quad \quad \quad \quad \quad \quad \quad \quad \quad  \quad \quad \quad \quad \quad
\begin{minipage}{0.2\linewidth} \label{fig2}
\begin{texdraw}
\setunitscale 0.3
\move(0 0) \lvec(1 0) \move(1 0.1) \lvec(1 1)
\move(1.1 1) \lvec(2 1) \move(2 1.1) \lvec(2 2)
\move(2.1 2) \lvec(3 2) \move(3 2.1) \lvec(3 3) \fcir f:0  r:0.08
\move(3.1 3) \lvec(4 3) \move(4 3.1) \lvec(4 4)
\move(4.1 4) \lvec(5 4) \move(5 4.1) \lvec(5 5)
\move(5.1 5) \lvec(6 5) \move(6 5.1) \lvec(6 6)

\end{texdraw}
\caption{Staircase initial value problems} \label{init2}
\end{minipage}
\end{figure}

\begin{figure}[h]
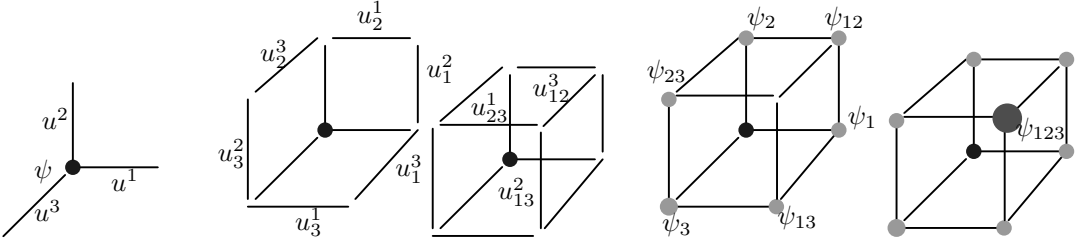

\begin{minipage}[t]{0.1\linewidth} \centering

\begin{texdraw}
\setunitscale 0.4

\move(0 0) \fcir f:0.1  r:0.1
\move (0.1 0) \lvec (1.1 0) 
\move (0 0.1) \lvec (0 1.1) 
\move (-0.1 -0.1) \lvec (-0.9 -0.9) 

\htext (-0.5  -0.2) {$ \psi$}
\htext (0.5  -0.3) {$u^1$}
\htext (-0.4  0.5) {$u^2$}
\htext (-0.5  -0.7) {$u^3$}
\end{texdraw}
\end{minipage}
\quad\quad\quad
\noindent \begin{minipage}[t]{0.1\linewidth} \centering

\begin{texdraw}
\setunitscale 0.4

\move(0 0) \fcir f:0.1  r:0.1
\move (0.1 0) \lvec (1.1 0) 
\move (0 0.1) \lvec (0 1.1) 
\move (-0.1 -0.1) \lvec (-0.9 -0.9) 
\move(1.2 0.1) \lvec(1.2 1.1) \move(0.1 1.2) \lvec(1.1 1.2) 
\move(-1 -1) \lvec(0.3 -1) \move(1.2 0) \lvec(0.39 -0.9) 
\move(-0.1 1.2) \lvec(-0.9 0.5) \move(-1. -0.9) \lvec(-1. 0.4)  

\htext (1.3  0.6) {$u^2_1$}   \htext (0.4  1.3) {$u^1_2$}
\htext (-0.4  -1.4) {$u^1_3$}   \htext (0.9  -0.7) {$u^3_1$}
\htext (-0.85  0.8) {$u^3_2$}   \htext (-1.4  -0.5) {$u^2_3$}
\end{texdraw}
\end{minipage}
\quad\quad\quad
\noindent \begin{minipage}[t]{0.1\linewidth}
\begin{texdraw}
\setunitscale 0.4
\move(0 0) \fcir f:0.1  r:0.1
\move (0.1 0) \lvec (1.1 0) 
\move (0 0.1) \lvec (0 1.1) 
\move (-0.1 -0.1) \lvec (-0.9 -0.9) 
\move(1.2 0.1) \lvec(1.2 1.1) \move(0.1 1.2) \lvec(1.1 1.2)       
\move(-1 -1) \lvec(0.3 -1) \move(1.2 0) \lvec(0.45 -0.9)         
\move(-0.1 1.2) \lvec(-0.9 0.5) \move(-1. -0.9) \lvec(-1. 0.4)        

\move(0.4 -0.9) \lvec(0.4 0.35)
\move(-0.9 0.45) \lvec(0.35 0.45)
\move(1.1 1.1)  \lvec(0.43 0.43)

\htext (0.3 0.75) {$u^3_{12}$}
\htext (-0.15 -0.6) {$u^2_{13}$}
\htext (-0.5 0.5) {$u^1_{23}$}
\end{texdraw}
\end{minipage}
\quad\quad\quad
\noindent \begin{minipage}[t]{0.1\linewidth}
\begin{texdraw}
\setunitscale 0.4
\move(0 0) \fcir f:0.1  r:0.1
\move (0.1 0) \lvec (1.1 0) 
\move (0 0.1) \lvec (0 1.1) 
\move (-0.1 -0.1) \lvec (-0.9 -0.9) 
\move(1.2 0.1) \lvec(1.2 1.1) \move(0.1 1.2) \lvec(1.1 1.2)       \move(1.2 1.2) \fcir f:0.6  r:0.1        \htext (1 1.3) {$\psi_{12}$}                 
\move(-1 -1) \lvec(0.3 -1) \move(1.2 0) \lvec(0.45 -0.9)         \move(0.39 -1) \fcir f:0.6  r:0.1         \htext (0.39 -1.3) {$\psi_{13}$}                 
\move(-0.1 1.2) \lvec(-0.9 0.5) \move(-1. -0.9) \lvec(-1. 0.4)          \move(-1 0.4) \fcir f:0.6  r:0.1    \htext (-1.3 0.6) {$\psi_{23}$}     

\move(0.4 -0.9) \lvec(0.4 0.35)
\move(-0.9 0.45) \lvec(0.35 0.45)
\move(1.1 1.1)  \lvec(0.43 0.43)

\move(1.2 0) \fcir f:0.6  r:0.1   \htext (1.3 0) {$\psi_{1}$}
\move(0 1.2) \fcir f:0.6  r:0.1    \htext (0 1.3) {$\psi_{2}$}
\move(-1. -1.) \fcir f:0.6  r:0.12   \htext (-1.1 -1.4) {$\psi_{3}$}
\end{texdraw}
\end{minipage}
\quad\quad\quad\quad
\noindent \begin{minipage}[t]{0.1\linewidth}
\begin{texdraw}
\setunitscale 0.4

\move(0 0) \fcir f:0.1  r:0.1
\move (0.1 0) \lvec (1.1 0) 
\move (0 0.1) \lvec (0 1.1) 
\move (-0.1 -0.1) \lvec (-0.9 -0.9) 
\move(1.2 0.1) \lvec(1.2 1.1) \move(0.1 1.2) \lvec(1.1 1.2)       \move(1.2 1.2) \fcir f:0.6  r:0.1              
\move(-1 -1) \lvec(0.3 -1) \move(1.2 0) \lvec(0.45 -0.9)         \move(0.39 -1) \fcir f:0.6  r:0.1                 
\move(-0.1 1.2) \lvec(-0.9 0.5) \move(-1. -0.9) \lvec(-1. 0.4)          \move(-1 0.4) \fcir f:0.6  r:0.1            

\move(0.4 -0.9) \lvec(0.4 0.35)
\move(-0.9 0.45) \lvec(0.35 0.45)
\move(1.1 1.1)  \lvec(0.43 0.43)      \move(0.43 0.43)          \fcir f:0.3  r:0.19  \htext (0.53 0.13) {$\psi_{123}$}

\move(1.2 0) \fcir f:0.6  r:0.1 
\move(0 1.2) \fcir f:0.6  r:0.1 
\move(-1. -1.) \fcir f:0.6  r:0.12 
\end{texdraw}
\end{minipage}
\caption{ The 3D-consistency property of the idolons.}
\label{evolut}
\end{figure}

In both cases one determines first the functions $u^i$ in the whole domain by means of \mref{supl} and then $\psi$ by means of \mref{impl} as illustrated on Figure \ref{evolut} in case of one cube.
 In the sequence of the sub-figures of Figure \ref{evolut} we demonstrate the 3D-consistency property of the idolons that correspond to the $I$ or the $cI$ lists. In the first sub-figure we specify the initial data, namely the value of the potential $\psi $  at the origin, together with the fields $u^i,\; i=1,2,3,$ (or $s^i,$ when we consider the $cI-$list ) at the 3 edges.  On the second  sub-figure, we are using the difference equations of Table \ref{IsF}, to calculate the fields $u^i_j$ (or $s^i_j$ of Table \ref{IsH}) $i,j=1,2,3,$    $i\neq j  \neq i$.  In the third sub-figure, we calculate    $u^i_{jk}$ (or $s^i_{jk}$),  $i,j,k=1,2,3,$    $i\neq j \neq k \neq i,$ taking advantage of the compatibility of the system (\ref{compa}).
 Then, in the fourth sub-figure by using the potential relations of   Table \ref{IsF} (or of Table \ref{IsH} when we deal with $cI-$list), fields $\psi_i$ are uniquely determined and fields $\psi_{ij}$ are specified in a unique way, see Table \ref{isF} (\ref{isH}).
 Finally, in the fifth sub-figure the value of $\psi_{ijk}$ can be computed threefold and each of the three different computations gives the same result.  Following exactly the same arguments, nD-consistency can be proven just by setting  $i,j,k=1,2,\ldots n,$    $i\neq j \neq k \neq i$.

\subsection{Equations from the ABS list}
\label{ABSs}
The problem of multivaluedness does not occur
when the function $f$ in formula \mref{ff} is fractional linear and the  potentials of systems presented in Section \ref{Systems} obey the  equations $Q(\psi_{12},\psi_1,\psi_2,\psi)=0$ where Q is affine-linear in each of its variables.
This case has been considered in details in \cite{Tasos}.
The  equations  we  obtain  are  point equivalent to the following equations from the ABS list
\begin{equation}
\begin{array}{ll}
H1: &  (\psi_{12}-\psi)(\psi_1-\psi_2) = \alpha-\beta,\cr
H2: & (\psi_{12}-\psi)(\psi_1-\psi_2) - (\alpha-\beta)(\psi_{12} + \psi_1 + \psi_2 + \psi)+(\alpha^2-\beta^2)=0 \cr
H3^{\delta}: &  \alpha(\psi_{12}\psi_2+\psi_1\psi)-\beta(\psi_{12}\psi_1+\psi_2\psi) +\delta (\alpha^2-\beta^2 ) = 0, \cr
A1^{\delta}:  &   \alpha(\psi_{12}+\psi_{1})(\psi_2 +\psi)-\beta(\psi_{12}+\psi_2)(\psi_1+\psi)  =  \delta^2\alpha \beta(\alpha-\beta) , \cr
A2: & (\alpha^2-\beta^2)(\psi_{12}\psi_1\psi_2\psi+1)+\alpha(\beta^2-1)(\psi_{12}\psi_2+\psi_1\psi)-\beta(\alpha^2-1)(\psi_{12}\psi_1+\psi_2\psi) = 0 , \cr
\end{array}
\end{equation}
where $\alpha$ and $\beta$ are given functions of $m^1$ and $m^2$ respectively, $\delta$ is a constant parameter that can be re-scaled without loss of generality to $\delta=0$ or $\delta=1$.
In some places we refer to equations
\begin{equation}
\begin{array}{ll}
Q1^{\delta} &  \alpha(\psi_{12}-\psi_{1})(\psi_2 -\psi)-\beta(\psi_{12}-\psi_2)(\psi_1-\psi) + \delta^2 \alpha\beta(\alpha-\beta) = 0, \cr
Q3^0: & (\alpha^2-\beta^2)(\psi_{12}\psi+\psi_1\psi_2)+\alpha(\beta^2-1)(\psi_{12}\psi_1+\psi_2\psi)-\beta(\alpha^2-1)(\psi_{12}\psi_2+\psi_1\psi) = 0 , \cr
\end{array}
\end{equation}
despite of the fact $Q1^{\delta}$ is point equivalent to $A1^{\delta}$ and $Q3^0$ is point equivalent to $A2$.
Conversely,  within ABS list only the above equations admit existence of the potentials $\psi$.  Mutual relations between the list of ABS equations and the bond systems are summarized in the Table \ref{e2m}.

\begin{table}[h!]
\caption{Relation of maps with the ABS list}
\label{e2m}
\begin{tabular}{llcr}
\toprule
 Equation & Relation to  map  & \parbox{3.6cm}{Parameter associations \\ \centerline{$(\alpha,\beta)$}} & Map \\ [3mm]\hline
\multirow{3}{*}{$H1$}      & $u=\psi_1+\psi$,~~~$v=\psi_2+\psi$ & $(p,q)$ & $F_{V}$   \\ [0mm]
                           & $u=\psi_1-\psi$,~$v=\psi_2-\psi$   & $(p,q)$ & $cH_{V}$  \\[0mm]
                           & $u=\psi_1\psi$,~$v=\psi_2\psi$     & $(p,q)$ & $F_{IV}$  \\ [0mm]
\hline
\multirow{1}{*}{$H2$}      & $2u=\psi_1+\psi-p$,~~~$2v=\psi_2+\psi-q$ & $(p,q)$& $F_{IV}$ \\ [0mm]
\hline
\multirow{3}{*}{$H3^{0}$}  & $\sqrt p u=\psi_1\psi$,~~~$\sqrt q v=\psi_2\psi$ & $(\sqrt p,\sqrt q)$& $F_{III}$ \\  [0mm]
                           & $\sqrt p u=(-1)^{m^1+m^2}\psi_1\psi$,~~~$\sqrt q v=(-1)^{m^1+m^2}\psi_2\psi$ & $(\sqrt p,\sqrt q)$& $cH^A_{III} $    \\ [0.5mm]
                           & $\sqrt p u=\frac{\psi_1}{\psi}$,~~~$\sqrt q v=\frac{\psi_2}{\psi}$ & $(1/\sqrt p,1/\sqrt q)$& $cH_{III}^B$   \\ [0.5mm]
\hline
\multirow{1}{*}{$H3^{-1}$} & $\sqrt p u=\psi_1\psi$,~~~$\sqrt q v=\psi_2\psi$ & $(\sqrt p,\sqrt q)$& $F_{II}$ \\ [0mm]
\hline
\multirow{1}{*}{$H3^{1}$}  & $\sqrt p u=-\psi_1\psi$,~~~$\sqrt q v=-\psi_2\psi$ & $(\sqrt p,\sqrt q)$& $F_{II}$ \\ [0mm]
\hline
\multirow{2}{*}{$Q1^{0}$} & $\frac{u}{u-1}=\frac{\psi_1}{\psi}$,~~~$\frac{v}{v-1}=\frac{\psi_2}{\psi}$ & $(p,q)$& $cH_{II}$ \\ [0.5mm]
                          & $pu=\psi_1-\psi$,~~~$qv=\psi_2-\psi$ & $(p,q)$& $cH^A_{III} $    \\ [0mm]
\hline $Q1^{\pm 1}$       & $2pu=-(\psi_1-\psi-p)$,~~~$2qv=-(\psi_2-\psi-q)$& $(p,q)$&$cH_{II}$     \\ [1mm]
\hline \multirow{2}{*}{$Q3^{0}$} &$\sqrt p u=-\frac{\psi_1}{\psi}$,~~~$\sqrt q v=-\frac{\psi_2}{\psi}$&$(-1/\sqrt p,-1/\sqrt q)$& $cH_I$\\ [1mm]
&$ \frac{u-p}{u-1}=\sqrt p \frac{\psi_1}{\psi}$,~~~$\frac{v-q}{v-1}=\sqrt q \frac{\psi_2}{\psi}$&$(\sqrt p,\sqrt q)$& $cH_I$\\ [0mm]
\hline
\multirow{1}{*}{
$A1^{\pm 1}$}
& $2pu- p=\psi_1+\psi$,~~~$2qv-q=\psi_2+\psi$ & $(p,q)$& $F_{II}$    \\ [0mm]
\hline
\multirow{3}{*}{
$A1^{0}$}
& $2pu=\psi_1+\psi$,~~~$2qv=\psi_2+\psi$ & $(p,q)$& $F_{III}$    \\ [0mm]
& $\frac{1}{u}=\psi_1+\psi$,~~~$\frac{1}{v}=\psi_2+\psi$ & $(p,q)$& $F_{III}$    \\ [0mm]
& $\frac{u}{u-1}=-\frac{\psi_1}{\psi}$,~~~$\frac{v}{v-1}=-\frac{\psi_2}{\psi}$ & $(p,q)$& $cH_{II}$ \\ [1mm]
\hline
  & $u=\sqrt p \psi_1\psi$,~~~$v=\sqrt q \psi_2\psi$ & $(\sqrt p,\sqrt q)$& $F_{I}$ \\  [0mm]
\multirow{3}{*}{$A2$} & $u=-\sqrt{p(p-1)}\psi_1\psi+p$,~~~$v=-\sqrt{q(q-1)}\psi_2\psi+q$ & $\left(\sqrt{\frac{p-1}{p}},\sqrt{\frac{q-1}{q}}\right)$& $F_{I}$   \\ [1mm]
  & $u=\sqrt{1-p}\psi_1\psi+1$,~~~$v=\sqrt{1-q}\psi_2\psi+1$ & $\left(-\frac{1}{\sqrt{1-p}},-\frac{1}{\sqrt{1-q}}\right)$& $F_{I}$ \\ [0mm]
\hline
\bottomrule
\end{tabular}
\end{table}

\begin{figure}[!h]
\begin{picture}(270,270)(-150,15)

\put(85,250){$I_I$}
\put(80, 240){\line(-1, -1){20}} \put(90, 265){\line(0, 1){20}} \put(100, 240){\line(1, -1){20}}
\put(43,200){$A2$}  \put(85,292){$A2 $}  \put(120,200){$A2$}

\put(-25,190){$I_{II}$}
\put(-30, 180){\line(-1, -1){20}} \put(-18, 205){\line(0, 1){20}} \put(-10, 180){\line(1, -1){20}}
\put(-65,140){$H3^{1}$}  \put(-35,232){$A1^{\pm 1}$}  \put(15,140){$H3^{-1}$}

\put(200,190){$I_{III}$}
\put(195, 180){\line(-1, -1){20}} \put(205, 205){\line(0, 1){20}} \put(215, 180){\line(1, -1){20}}
\put(155,140){$A1^{0}$}  \put(195,232){$H3^0$}  \put(240,140){$A1^{0}$}
\put(45,80){$I_V$}  \put(125,80){$I_{IV}$}
\put(40, 70){\line(-1, -1){20}} \put(52, 95){\line(0, 1){20}} \put(60, 70){\line(1, -1){20}}
\put(120, 70){\line(-1, -1){20}}
\put(132, 95){\line(0, 1){20}}
\put(140, 70){\line(1, -1){20}}
\put(3,30){$H2^*$}  \put(45,122){$F1$}  \put(83,30){$H1$}
\put(125,122){$G2$} \put(162,30){$H2$}
\end{picture}
\caption{Systems and their spanning idolons}
\label{pic}
\end{figure}
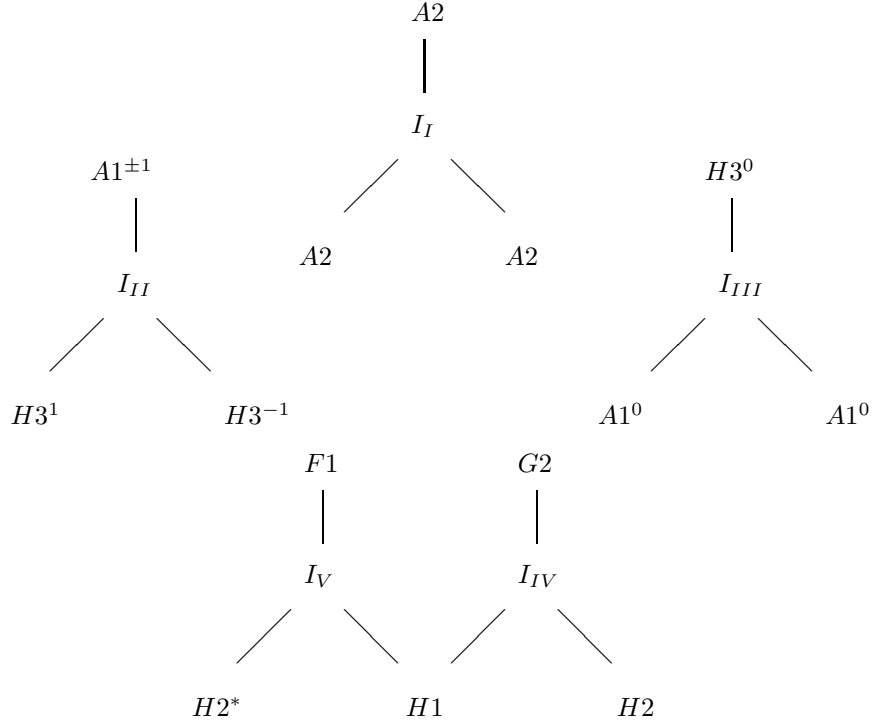

Any linear combination of potentials is potential again. We span the space of potentials in a basis. As a natural basis in the case of the
systems $I_I$, $I_{II}$, $I_{III}$ can be chosen the basis consisting of potentials that are governed by  multilinear equations see Figure \ref{pic}.
However, in the case of systems $I_V$, $I_{IV}$ one has to take into consideration the higher degree equations $H2^*$, $G2$ and $F1$. $H2^*$ is a multi-quadratic quad-relation that was first obtained in \cite{AdVeQ}, see also \cite{KaNie,AtkNie}. $G2$ is a multi-quartic quad relation that was first presented implicitly in \cite{KaNie}. $F1$ is quad-relation which the variables appear of degree  nine, it was also presented implicitly \cite{KaNie}. The last Figure of this paper reveals intriguing structure of the difference systems and their potentials with the top model  spanned by three $A2$ equations  (it turns out it can be decomposed into three versions of multi-quadratic $A2^*$ relations of the papers \cite{JJJ,AtkNie}), with the structure of the models $I_{II}$ and $I_{III}$ which resembles the quark structure ``neutron-proton" and with the non-multilinearity  even in models forming the basis of the bottom models $I_{IV}$ and $I_V$.

\section{B\"acklund transformations and perspectives} \label{Section7}
Despite all the idolons are consistent-around-the-cube which is itself regarded as hallmark of integrability, we see their  integrability somewhere else. Namely, in each family of the systems of Tables \ref{IsF}, \ref{isH} there is at least one idolon which is multilinear. There exists a map (non-auto B\"acklund transformation) that transform the solutions of the multilinear equation to solutions of  any idolon of the family.  For instance,  in the case of the system $I_I$
 a solution $x$ of  $A2$ equation define functions $u^i=x x_i.$ Pluging it into the second expression from the first row of Table \ref{IsF} we get ($y:=e^\psi$)
\[
y_iy  = (x x_i)^a\left( \frac{x x_i-p^i}{\sqrt{{p^i}^2-1}}\right)^b  \left(\frac{p^ix x_i-1}{\sqrt{{p^i}^2-1}}\right)^c
\]
which constitutes B\"acklund transformation from equation $A2$ to the following system
\begin{equation}
\begin{array}{l}
y_iy  = (u^i)^a\left( \frac{u^i-p^i}{\sqrt{{p^i}^2-1}}\right)^b  \left(\frac{p^iu^i-1}{\sqrt{{p^i}^2-1}}\right)^c, \\
\frac{y_{ij}}{y}=\left( \frac{p^i [1-{p^j}^2] u^i - p^j [1-{p^i}^2] u^j-[{p^i}^2-{p^j}^2]}{p^j [1-{p^i}^2] u^i - p^i [1-{p^j}^2] u^j+({p^i}^2-{p^j}^2)u^iu^j}\right)^a
 \left(\frac{(p^ju^j-p^iu^i)\sqrt{{p^i}^2-1}\sqrt{{p^j}^2-1}}{p^j [1-{p^i}^2] u^i - p^i [1-{p^j}^2] u^j+({p^i}^2-{p^j}^2)u^iu^j}\right)^b \cdot\\
\qquad  \qquad \qquad\qquad\cdot \left( \frac{(p^iu^j-p^ju^i)\sqrt{{p^i}^2-1}\sqrt{{p^j}^2-1}}{p^j [1-{p^i}^2] u^i - p^i [1-{p^j}^2] u^j+({p^i}^2-{p^j}^2)u^iu^j} \right)^c.
\end{array}
\end{equation}
In this way we got an extension of results of the papers \cite{James,Boll} where  systematic  studies on B\"acklund tranformations between equations from ABS list can be found. However,
we concentrated here on non-auto B\"acklund transformations, which map solutions $x$ of an equation of the ABS list to solutions $y$ of another equation, which can ce reduced to the following
form
\begin{equation}
\label{type}
y_i\pm y=u(x,x_i),
\end{equation}
or to the exponentiated form
\begin{equation}
\label{typee}
y_iy^{\pm 1}=u(x,x_i).
\end{equation}
For a given equation we gave a procedure that allows one to find all the equations that can be connected to the original equation by B\"acklund transformations of type  \mref{type} or of its exponentiated form \mref{typee}.
It is urge to extend this procedure to more complex but still linearizable transformations of the form
\begin{equation}
y_i=\frac{s^i(x,x_i)y+u^i(x,x_i)}{t^i(x,x_i)y+v^i(x,x_i)}, \qquad i=1,\ldots,n.
\end{equation}


\begin{thebibliography}{10}

\bibitem{ABS}
V.E. Adler, A.I. Bobenko, and Yu.B. Suris.
\newblock Classification of integrable equations on quad-graphs. {T}he
  consistency approach.
\newblock {\em Comm. Math. Phys.}, 233(3):513--543, 2003.

\bibitem{ABSf}
V.E. Adler, A.I. Bobenko, and Yu.B. Suris.
\newblock Geometry of {Y}ang-{B}axter maps: pencils of conics and
  quadrirational mappings.
\newblock {\em Comm. Anal. Geom.}, 12(5):967--1007, 2004.

\bibitem{AdVeQ}
V.E. Adler and A.P. Veselov.
\newblock Cauchy problem for integrable discrete equations on quad-graphs.
\newblock {\em Acta Appl. Math.}, 84(2):237--262, 2004.

\bibitem{James}
J.~Atkinson.
\newblock {B}{\"a}cklund transformations for integrable lattice equations.
\newblock {\em J. Phys. A: Math. Theor.}, 41:Art. no. 135202, 2008.

\bibitem{JJJ}
J.~Atkinson.
\newblock {A} multidimensionally consistent version of {H}irota's discrete
  {K}d{V} equation.
\newblock {\em J. Phys. A: Math. Theor.}, 45:Art. no. 222001, 2012.

\bibitem{AtkNie}
J.~Atkinson and M.~Nieszporski.
\newblock Multi-quadratic quad equations: integrable cases from a
  factorised-discriminant hypothesis.
\newblock {\em Int. Math. Res. Not.}, 2220(15):4215--4240, 2013.

\bibitem{Boll}
R.~Boll.
\newblock Classification of 3{D} consistent quad-equations.
\newblock {\em J. Nonlinear Math. Phys.}, 18(3):337--365, 2011.

\bibitem{Dimakis2017}
A.~Dimakis and F.~M{\"u}ller-Hoissen.
\newblock Matrix {K}{P}: tropical limit and {Y}ang-{B}axter maps.
\newblock {\em Lett. Math. Phys.}, 2018.

\bibitem{Franks-book}
N~Hietarinta, J.~Joshi and F.W. Nijhoff.
\newblock {\em Discrete Systems and Integrablity}.
\newblock Cambridge Texts in Applied Mathematics (No. 54). Cambridge University
  Press, 2016.

\bibitem{Korepanov-1998}
R.M. Kashaev, I.G. Korepanov, and S.M. Sergeev.
\newblock Functional tetrahedron equation.
\newblock {\em Theor. Math. Phys.}, 117:1402--1413, 1998.

\bibitem{KaNie:2011}
P.~Kassotakis and M.~Nieszporski.
\newblock Families of integrable equations.
\newblock {\em SIGMA}, 7:100, 2011.

\bibitem{KaNie}
P.~Kassotakis and M.~Nieszporski.
\newblock On non-multiaffine consistent-around-the-cube lattice equations.
\newblock {\em Phys. Lett. A}, 376(45):3135--3140, 2012.
\newblock arXiv:1106.0435.

\bibitem{KaNie:2017}
P.~Kassotakis and M.~Nieszporski.
\newblock $2^n-$rational maps.
\newblock {\em J. Phys. A: Math. Theor.}, 50(21):21LT01, 2017.

\bibitem{KaNie:2018}
P.~Kassotakis and M.~Nieszporski.
\newblock Difference systems in bond and face variables and non-potential
  versions of discrete integrable systems.
\newblock {\em J. Phys. A: Math. Theor.}, 51(38):385203, 2018.

\bibitem{Kouloukas-2017}
T.~Kouloukas.
\newblock Relativistic collisions as {Y}ang-{B}axter maps.
\newblock {\em Phys. Lett. A}, 381(40):3445 -- 3449, 2017.

\bibitem{Nijhoff-1995}
F.~Nijhoff and H.~Capel.
\newblock The discrete korteweg-de vries equation.
\newblock {\em Acta Applicandae Mathematica}, 39(1):133--158, 1995.

\bibitem{nij-qui-cap}
F.W. Nijhoff, G.R.W. Quispel, and H.W. Capel.
\newblock Direct linearization of nonlinear difference-difference equations.
\newblock {\em Phys. Lett. A}, 97:125--128, 1983.

\bibitem{PSTV}
V.G. Papageorgiou, Yu.B. Suris, A.G. Tongas, and A.P. Veselov.
\newblock On quadrirational {Y}ang-{B}axter maps.
\newblock {\em SIGMA}, 6:9pp, 2010.

\bibitem{Tasos}
V.G. Papageorgiou, A.G. Tongas, and A.P. Veselov.
\newblock Yang-{B}axter maps and symmetries of integrable equations on
  quad-graphs.
\newblock {\em J. Math. Phys.}, 47:Art. no. 083502, 2006.

\bibitem{Sergeev-1998}
S.M. Sergeev.
\newblock Solutions of the functional tetrahedron equation connected with the
  local {Y}ang-{B}axter equation for the ferro-electric condition.
\newblock {\em Lett. Math. Phys.}, 45(2):113--119, 1998.

\bibitem{VeselovYB}
A.P. Veselov.
\newblock Yang-{B}axter maps and integrable dynamics.
\newblock {\em Phys. Lett. A}, 314:214--221, 2003.

\end{thebibliography}


\end{document}